\documentclass[pre,superscriptaddress,showpacs,twocolumn]{revtex4}
\usepackage{graphics}

\begin{document}

\title{An indicator for community structure}

\author{V. Gol'dshtein}
\email{vladimir@bgu.ac.il} \affiliation{Department of Mathematics,
Ben Gurion University of the Negev, Beer Sheva 84105, Israel}

\author{G.A. Koganov}
\email{quant@bgu.ac.il} \affiliation{Physics Department, Ben
Gurion University of the Negev, P.O.Box 653, Beer Sheva 84105,
Israel}

\begin{abstract}
An indicator for presence of community structure in networks is
suggested. It allows one to check whether such structures can
exist, in principle, in any particular network, without a need to
apply computationally cost algorithms. In this way we exclude a
large class of networks that do not possess any community
structure.
\end{abstract}


\maketitle

\section*{Introduction}

Community structure of networks have been intensively studied in
recent years, and a number of algorithms for finding such
structures have been suggested \cite{Newman-Girvan,review}. Being
quite effective, these algorithms allow one to find the community
structure in a wide variety of networks. However, such algorithms
are quite complicated and computationally demanding, moreover not
all networks possess any community structure at all. Therefore it
seems desirable to have some simple enough indicator allowing one
to judge about potential existence of community structures in
networks prior to exploiting the algorithms for finding them. The
usage of such an indicator could be quite effective in negative
sense, i.e. the negative answer to the question whether a
particular network can have any community structure or not would
allow one to exclude such a network from consideration, thus
avoiding a use of complicated numerical procedures for community
structure evaluation.

In this paper we propose an indicator of existence of community
structure in networks, which is based on a geometrically motivated
measure that compares an average network distance with its "mean
diameter". The indicator is oriented on relatively dense
communities, which is typical for sociological type of networks
\cite{social}. We provide some asymptotic estimations for this
indicator which allow one to evaluate its numerical values. The
indicator is applied to some model networks with dense
communities. Some real networks are analyzed as well.

\section*{Dilatation of a network as an indicator for community structure}

Our goal is to find relatively simple indicator for community
structure existence. The idea is to compare the mean distance,
which can be easily calculated, with some etalon characteristic of
networks of a given size $N$ and a mean degree $k$. To define such
an etalon let us look at some geometric analogy. Consider an
$n$-dimensional geometric body of volume $V$. In geometry the
ratio between the diameter of the body to its minimal possible value (which
is of order of $\sqrt[n]{V}$) is known as a dilatation
\cite{Rickman}, which is a measure of body's asymmetry.

We adopt this concept to networks using the notion of a mean distance $\bar{L}$ instead of the diameter. First, notice that the size $N$ of
a network can serve as an analog of geometric volume, and the
ratio $\delta$ of the number of links to the number of nodes $N$
can be chosen as an analog of dimension. For undirected networks
considered here, the dimension analog is $\delta=k/2$, where $k$
is the mean degree. For instance, the mean degree of a 2D-torus is
$k=4$, for 3D-torus $k=6$, etc.

We introduce a notion of dilatation of a network and define it as

\begin{equation}
D=\frac{\bar{L}}{\sqrt[\delta]{N}}.\label{Dilatation}
\end{equation}

\noindent Intuitively, a large value of dilatation can be caused
by strong inhomogeneity, which in turn can be a consequence of the
presence of community structure in the network. However, the
presence of community structure is just one of possible reasons
for high network dilatation. For instance, highly stretched
networks, such as a narrow long strip, can also have high value of
dilatation. Therefore the dilatation can only \textit{indicate} to
possible existence of a community structure. In other words, big
value of dilatation can serve as a necessary, but not sufficient
condition for community structure existence.

To illustrate the relationship between community structure and
dilatation consider a simplest possible configuration having an
evident community structure, namely, a network consisting of two
complete graphs of size $N/2$ each connected by a single link to
each other. This network contains two communities of maximal
density with a single inter-community link. For large networks ($N\gg 1)$ the mean distance and the dimension are $\bar{L}\approx 2$ and
$\delta\approx N/4$, respectively, so the dilatation is (see Appendix for details)

\begin{equation}
D\approx \frac{2}{\sqrt[N/4]{N}}\rightarrow 2. \label{twoGraphs}
\end{equation}

As another extreme example consider a network without any
community structure, namely a binary tree of $M$ levels. The
network size is $N=2^{M+1}-1$, the total number of links equals to
$N-1$, so the dimension $\delta \approx 1$, and the mean distance
can be estimated as $L\approx 2M$. Substituting $L$ and $\delta$
into Eq.(\ref{Dilatation}) results in the following asymptotic
expression for dilatation:

\begin{equation}
D\approx \frac{log_{2}N}{N}\rightarrow 0. \label{tree}
\end{equation}

The described extreme examples prompt us to make a plausible
conjecture that, in general, the dilatation of networks baring no
community structure should be relatively small (less than one),
whereas the dilatation of well-structured networks should be at
least around 2. Below the last statement will be corrected based
on both analytical estimations and further examples.

\subsection*{Asymptotical estimations}

To develop analytical estimates we need to describe more formally
community networks. Consider a global connected network $G$
comprising of $N$ nodes and $E$ links, so that the dimension of
$G$ is $\delta=E/N$. To estimate the dilatation we need, according
to Eq.(\ref{Dilatation}), an estimation for the mean distance
$\bar{L}_{G}$. To calculate $\bar{L}_{G}$ suppose that the network is
divided into $C$ communities, so that $i-th$ community contains
$m_{i}$ nodes. Let us also assume that any two communities are
connected by a single link at maximum. It should be noted that
being obtained under this assumption, our analytical estimations
work quite well also beyond it, as will be demonstrated below. Let
us define a macro network $M$ by replacing each of $C$ communities
with a single node and denote the mean distance in $M$ by
$\bar{L}_{M}$. The formulas for the dilatation and the mean
distance in general case, derived in Appendix, assume some apriory
knowledge about the structure of the network. It is possible
however to get, under some additional assumptions, analytical
estimations which do not demand any apriory knowledge about the
structure of the network. Indeed, assuming that all communities
have the same size $m_{i}=m\gg 1$ and the same mean distance $\bar{L}_{C}$, we obtain the following
estimation for the mean distance $\bar{L}_{G}$ of the global network

\begin{equation}
\overline{L}_{G}\approx\left[\frac{1}{C}+\frac{2(C-1)}{C}\right]\overline{L}_{C}+\frac{C-1}{C}\overline{L}_{M},
\label{MeanDistance}
\end{equation}

\noindent where $\overline{L}_{C}$ is the mean distance inside a
single community. Equation (\ref{MeanDistance}) still bears some
apriory information about the community structure, namely the
number of communities $C$ and the mean distance $\overline{L}_{C}$
inside a single community. Assuming that the number of communities
is large enough $C \gg 1$ and taking into account that $\bar{L}_{C}\geq
1$ (equality takes place for a complete graph), results in

\begin{equation}
\overline{L}_{G}\geq \overline{L}_{M}+2. \label{RoughMeanDistance}
\end{equation}

Inequality (\ref{RoughMeanDistance}) means that given a
particular network with the mean distance $\overline{L}_{G}$, any
reasonable partition of the network into dense communities will
result in a macro network whose mean distance is not bigger than
$\overline{L}_{G}-2$. If, on the other hand, after partitioning the
network inequality (\ref{RoughMeanDistance}) is violated, then the
partition was not suitable. Notice that equality in
(\ref{RoughMeanDistance}) takes place when all communities are
complete graphs (1-cliques \cite{social}).

Using Eq.(\ref{MeanDistance}) results in the following estimation for dilatation:

\begin{equation}
D\approx
N^{-1/\delta}\left(\left[\frac{1}{C}+\frac{2(C-1)}{C}\right]\overline{L}_{C}+\frac{C-1}{C}\overline{L}_{M}\right).
\label{DilAnalytic}
\end{equation}

Taking into account that $C \geq 2$, $\bar{L}_{C}\geq 1$ and $\bar{L}_{M}\geq
1$ (equality takes place for a complete graph) we obtain  from
Eq.(\ref{DilAnalytic}) the following rough estimate which does not
depend on any apriory knowledge about the community structure:

\begin{equation}
D \geq \frac{2}{\sqrt[\delta]{N}}. \label{RoughDilAnalytic}
\end{equation}

For the case  $C \gg 1$ it follows from
Eqs.(\ref{RoughMeanDistance}) and (\ref{DilAnalytic}) that

\begin{equation}
D\geq \frac{\overline{L}_{M}+2}{\sqrt[\delta]{N}}\geq
\frac{3}{\sqrt[\delta]{N}}. \label{RoughAsympDilatation}
\end{equation}

Thus the behavior of the dilatation depends on the parameter
$N^{1/\delta}=N^{N/E}$ that we call a mean diameter of the network.

This study is focused on Sparse networks with Dense communities
(SD-networks), in which the value of $N^{1/\delta}$ is close to
unity. We propose the following formal definition of SD-networks:
A network $G$ is an SD-network if $\sqrt[\delta]{N}\rightarrow 1$
for $N\rightarrow\infty$. To satisfy this condition it is enough
that the dependence of $\delta = E/N$ upon $N$ has the form of
$E\geq N^s$ for $s>1$. As it follows from the above definition,
for any large ($N \gg 1$) SD-network $D \geq 2$, where the
estimation (\ref{RoughMeanDistance}) was used. We remind that
according to eq.(\ref{twoGraphs}) for the network consisting of
two complete graphs of size $N/2$ connected by a single link the
dilatation $D\approx 2/\sqrt[N/4]{N}\rightarrow 2$.

It should be noted that the asymptotic estimation $D \geq 2$ was
obtained under additional restrictions on networks, namely: (i)
all communities are large enough and have approximately the same
size, (ii) all communities have the same mean distance, and (iii)
there exists not more then one link between two different
communities. However, in real-life networks these restrictions can
be violated, therefore we suggest to use $D>1$ as a criterium for
community structure existence. Numerical simulations presented in
the next section support the choice of the criterium $D>1$.

\section*{Model examples}

In this section we demonstrate, using numerical simulations, how
the introduced above indicator of existence of community structure
works on some model networks. It will be shown that the criterium
$D>1$ is quite reasonable, even for relatively small communities.
Also, it will be demonstrated that being derived in asymptotical
limit $m\gg 1, C\gg 1$, analytical estimations
Eqs.(\ref{RoughMeanDistance}) and (\ref{RoughDilAnalytic}) are not
restricted by this limit, and work quite well even for relatively
small values of $m$ and $C$.

\begin{figure}[ht]
\centerline{\scalebox{0.9}{\includegraphics{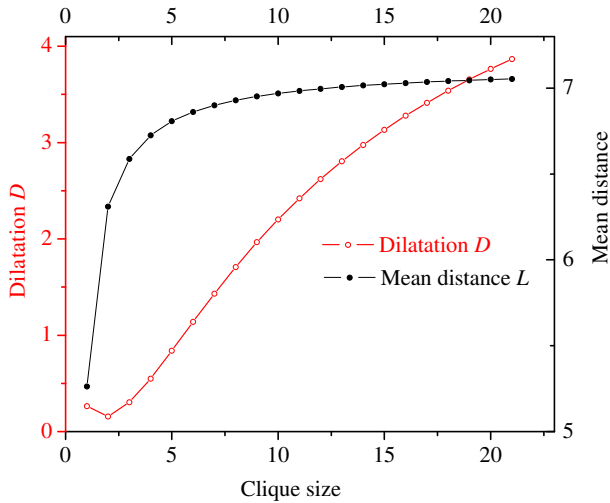}}}
\caption{\label{Ring20-2} Ring network consisting of 20
complete-graph-clusters, $k=2$. Dependence of dilatation mean
distance upon cluster size.}
\end{figure}
\begin{figure}[ht]
\centerline{\scalebox{0.9}{\includegraphics{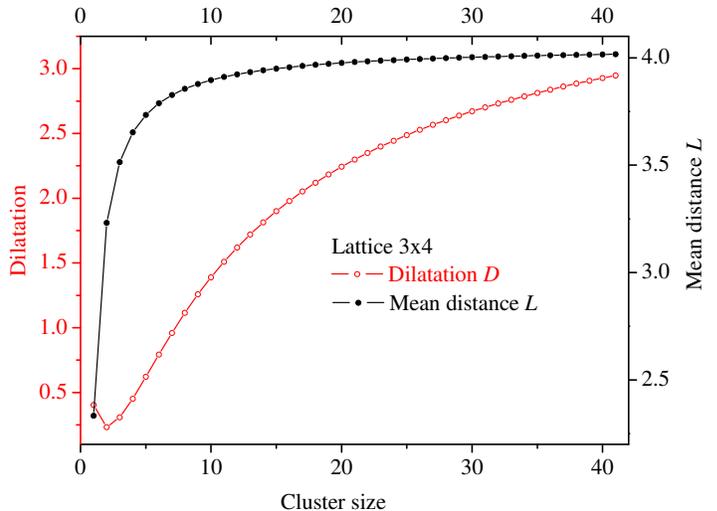}}}
\caption{\label{Lattice3x4} 3x4 lattice. Dependence of dilatation
and the mean distance on cluster size.}
\end{figure}

As a first model example consider a ring network with $n=20$ and
$k=2$, in which each of all 20 nodes is replaced with a complete
graph of size $m$. Two quantities have been calculated, the
dilatation $D$, and the mean distance $\overline{L}_{G}$ as functions
of $m$. The result is shown in Fig. \ref{Ring20-2}. One can see
that the for initial simple ring with $m=1$, that has no any
structure, the dilatation is small. While the community size $m$
increases, the dilatation grows and exceeds 1 at $m>5$, when the
community structure becomes clearly pronounced. One can also
notice that starting from $m\leq 10$ the analytical estimations
Eqs.(\ref{RoughMeanDistance}) and (\ref{RoughDilAnalytic}) hold
well. The same comment holds true for the lattice network (see
Fig.\ref{Lattice3x4}).

Both the analytical estimations Eqs.(\ref{RoughMeanDistance}) and
(\ref{RoughDilAnalytic}) and the results shown in Figs.
\ref{Ring20-2} and \ref{Lattice3x4}, have been obtained under
assumption that all communities are of the same size. However,
applicability of the estimations Eqs.(\ref{RoughMeanDistance}) and
(\ref{RoughDilAnalytic}) is not restricted to this assumption. To
demonstrate this, we have constructed a network with variable
community size, namely 4x3 lattice with community sizes randomly
chosen between 20 and 60, so that the mean community size equals
to 40.  Calculation of the mean distance and the dilatation of
this network gives $D=2.98 \pm 0.07$ and $L=3.98 \pm 0.06$. The
same network with equal community sizes corresponds to the last
point in Fig.\ref{Lattice3x4} where the dilatation and the mean
distance are $D=2.92782$ and $L=4.01601$, correspondingly.
Comparing the dilatation and the mean distance for these two
networks one can conclude that the estimations
Eqs.(\ref{RoughMeanDistance}) and (\ref{RoughDilAnalytic}) work
well for networks with not equal communities as well.

\begin{figure}[ht]
\centerline{\scalebox{0.9}{\includegraphics{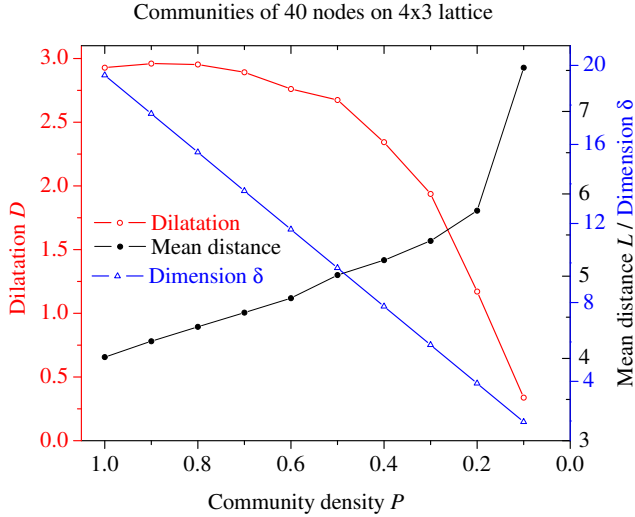}}}
\caption{\label{noncomplete} 3x4 lattice. Dependence of dilatation
and the mean distance on the number of inside-community links.}
\end{figure}

Consider again the 4x3 lattice network with equal communities, but
now the communities are not complete graphs, namely each pair of nodes inside the communities is linked with probability $P$ called community density. So the number of
inside-community links varies. The dilatation and the mean distance as functions of the community density $P$ is shown in Fig.\ref{noncomplete}.

As it can be seen from Fig.\ref{noncomplete}, the dilatation keeps above unity even at quite low community density ($P>0.2$), indicating to existense of community structure. Moreover, the estimations
Eqs.(\ref{RoughMeanDistance}) and (\ref{RoughDilAnalytic}) work
quite well, as long as the density of communities is not too low. At low values of $P<0.2$ the dilatation is less than 1 indicating to the absence of dense communities. This is consistent with the fact that the network dimension becomes low as well. We note again that both our indicator and analytical estimations are applicable to SD-networks (relatively dense communities sparsely connected to each other).

\begin{figure}[ht]
\centerline{\scalebox{0.9}{\includegraphics{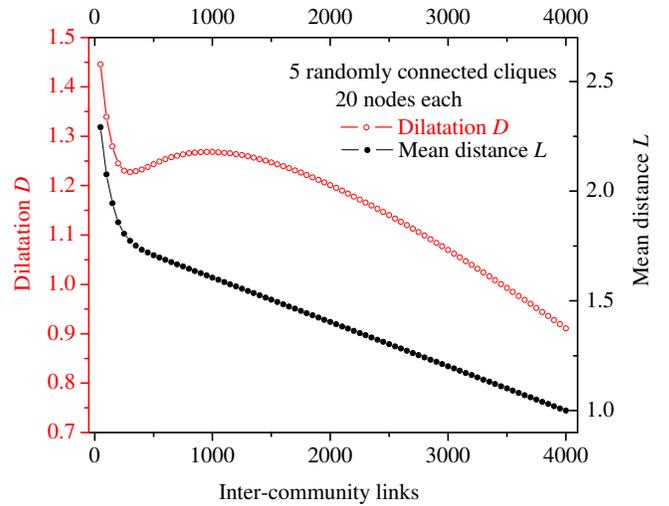}}}
\caption{\label{Cliques5x20} 5 randomly connected cliques, 20
nodes each. Dependence of the dilatation and the mean distance
upon the number of inter-community links.}
\end{figure}

Another assumption made in the course of derivation of analytical
estimations Eqs.(\ref{RoughMeanDistance}) and
(\ref{RoughDilAnalytic}) is that each pair of communities is
connected by not more than one link. To check what happens when we
go beyond this assumption, consider a network comprising of 5
randomly connected complete graphs, with the number of
inter-community links increased gradually from the minimal value
of 4 to the maximum of about 4000. Figure \ref{Cliques5x20} shows
the dependence of the mean distance and of the dilatation upon the
number of inter-community links. As one can see, the dilatation
remains above 1 up to quite large number of inter-community links
(about 2000), afterwards different communities become overlapped,
so that the border between communities cannot be defined clearly
enough. At the point when the number of inter-community links
reaches its maximum (4000) and $\delta = 50$, the entire network
becomes a complete graph, so that the mean distance equals to 1
and the dilatation is about 0.91. This value of dilatation differs
from asymptotical one for complete graph due to the fact that the
asymptotic $\sqrt[\delta]{N}\rightarrow 1$  is quite slow with
respect to $\delta$, therefore $\sqrt[50]{N}=1.096$.

\begin{figure}[ht]
\centerline{\scalebox{0.9}{\includegraphics{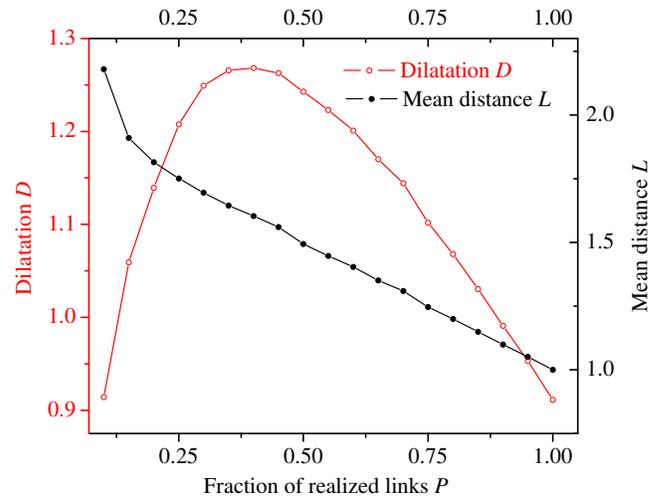}}}
\caption{\label{Random} Random graph of 100 nodes. Dilatation and
mean distance as functions of graph connectivity $P$.}
\end{figure}

\subsection*{Random graph}
Recently Guimer\'{a} et al. \cite{Guimera} have pointed out that
Rene-Erdos random graph \cite{Erdos} can exhibit a community
structure due to fluctuations. Their observation was based on a
concept of modularity introduced in Ref.\cite{Newman-Girvan}. In
this context it seems interesting to analyze such random graphs
from the point of view of dilatation. To do this we have
constructed a random graph containing 100 nodes connected to each
other with probability $P$, and calculated its dilatation. In Fig.
\ref{Random} both the dilatation and the mean distance are plotted
as functions of $P$. One can notice a clear maximum at $P=0.4$
where the dilatation, $D\approx 2.4$. According to our concept
this high value of dilatation indicates to possible existence of a
community structure in the network. This conclusion is consistent
with high modularity of random graphs reported in Ref.
\cite{Guimera}. It also seems that this can be related to another observation about possible hierarchy structure in random graphs \cite{our-hierarchy}.

\subsection*{Real-life networks}

To check how our indicator of the existence of community
structures works for real-life networks, we have calculated the
dilatation for 13 undirected networks using the data presented in the
Table II from Ref. \cite{Newman-SIAM}. The highest value of
dilatation $D=2.766$ was obtained for the network of film actors
\cite{actors}, which indicates that this network should be well
structured. Three other networks, train roots \cite{trainRouts},
Internet \cite{Internet}, and company directors
\cite{CompanyDirectors}, can also have community structure
according to their dilatation values of 1.78, 1.48, and 1.33,
respectively. All other networks presented in the table II from
Ref. \cite{Newman-SIAM} have the dilatation less than 1, therefore
according to criterium $D>1$ they hardly can have a community
structure. However, one should be careful using this criterium and
keep in mind the assumptions under which it was obtained. For
instance, looking at the data presented in the above mentioned
table we notice that despite the fact that some networks, like
math coauthorship, peer-to-peer and some other networks, have low
dilatation, they are not dense enough (not SD-networks) to apply
our criterium.

\section*{Summary}

The notion of dilatation of networks has been introduced.
Analytical estimations for the dilatation have been obtained under
some reasonable assumptions. The value of dilatation is suggested
to use as an indicator of existence of community structure in
Sparse networks with Dense communities (SD-networks). Both some
model and real-life networks have been considered to illustrate
the usage of the indicator suggested, as well as the applicability
of the analytical estimations. Numerical simulations
demonstrate that the analytical estimations can also be useful
beyond the assumptions made during the derivation.

\section*{Appendix}

This appendix contains analytical estimates of mean distances and
dilatations of SD-networks. All estimates are obtained assuming existence of
not more then one external link between different communities.
This restriction allows one to obtain comparatively simple and
compact estimates.

\subsection{Mean distance of networks with community structure}

This section is devoted to analytical estimates for mean distances
of SD-networks.

\subsubsection*{Mean distance estimates for networks consisting of two communities}

Consider two communities $Q_{1}$ and $Q_{2}$ containing $m_{1}$ and
$m_{2}$ nodes respectively, connected by a path
$\gamma$ of length $\left|\gamma\right|$. The path $\gamma$ connects an external
node $x_{0}$ belonging to the first community $Q_{1}$ with another
external node $y_{0}$ belonging to the second community $Q_{2}$.
The global network $G:=Q_{1}\cup Q_{2}\cup\gamma$ represents the
simplest possible example of a network with a community structure.

Introduce the following notation: $l(x_{i},x_{j})$ is the shortest
distance between nodes $x_{i},x_{j}$ of the first community
$Q_{1}$; $l(y_{s},y_{k})$ is the shortest distance between nodes
$y_{s},y_{k}$ of the second community $Q_{2}$. Hence the mean
distance $\overline{L}_{1,0}$ between the external node $x_{0}$
and other nodes of the community $Q_{1}$ is equal to
$\frac{1}{m_{1}-1}\sum_{i=1}^{m_{1}}l(x_{i},x_{0})$. By the
similar way define the mean distance
$\overline{L}_{2,0}=\frac{1}{m_{2}-1}\sum_{s=1}^{m_{2}}l(y_{s},y_{0})$
between the external node $y_{0}$ and other nodes of the community
$Q_{2}$ . Denote by
$\overline{L}_{1}=\frac{2}{m_{1}(m_{1}-1)}\sum_{i>j=1}^{m_{1}}l(x_{i},x_{j})$
the mean distance of the community $Q_{1}$ and by
$\overline{L}_{2}=\frac{2}{m_{2}(m_{2}-1)}\sum_{s>k=1}^{m_{2}}l(y_{s},y_{k})$
the mean distance of the community $Q_{2}$.

The mean distance for the global network $G$ can be calculated as

\begin{widetext}

\begin{equation}\label{LG-full}
\overline{L}_{G}=\frac{2\left(\sum_{i>j=1}^{m_{1}}l(x_{i},x_{j})+\sum_{s>k=1}^{m_{2}}l(y_{s},y_{k})\right)}{(m_{1}+m_{2})(m_{1}+m_{2}-1)}+\frac{2\sum_{i=1}^{m_{1}}\sum_{s=1}^{m_{2}}(l(x_{i},x_{0})+l(y_{s},y_{0})+\left|\gamma\right|)}{(m_{1}+m_{2})(m_{1}+m_{2}-1)}.
\end{equation}

Denote the first term of this sum as $I_{1}$ and the second term
as $I_{2}$. Using quantities $\overline{L}_{1},\overline{L}_{2},
\overline{L}_{1,0}$, and $\overline{L}_{2,0}$ the terms $I_{1}$
and $I_{2}$ can be written in a more compact way:

\begin{eqnarray}
I_{1}=\frac{m_{1}(m_{1}-1)\overline{L}_{1}+m_{2}(m_{2}-1)\overline{L}_{2}}{(m_{1}+m_{2})(m_{1}+m_{2}-1)},
\label{I1full}\\
I_{2}=\frac{2m_{2}(m_{1}-1)\overline{L}_{1,0}+2m_{1}(m_{2}-1)\overline{L}_{2,0}+2m_{1}m_{2}\left|\gamma\right|}{(m_{1}+m_{2})(m_{1}+m_{2}-1)}.\label{I2full}
\end{eqnarray}

Then

\begin{equation}\label{LG2}
\overline{L}_{G}=I_{1}+I_{2}=\frac{m_{1}(m_{1}-1)\overline{L}_{1}+m_{2}(m_{2}-1)\overline{L}_{2}}{(m_{1}+m_{2})(m_{1}+m_{2}-1)}+
\frac{2m_{2}(m_{1}-1)\overline{L}_{1,0}+2m_{1}(m_{2}-1)\overline{L}_{2,0}+2m_{1}m_{2}\left|\gamma\right|}{(m_{1}+m_{2})(m_{1}+m_{2}-1)}.
\end{equation}

\end{widetext}

The first term $I_{1}$ depends only on the mean distances
$\overline{L}_{1}$ and $\overline{L}_{2}$ inside the communities
$Q_{1}$ and $Q_{2}$ respectively, while the second term $I_{2}$
depends on the inter-community structure.

For big communities, when $m_{1},m_{2}\gg1$, Eq.(\ref{LG2}) for
the mean distance $L_{G}$ of the global network takes the
following asymptotic form

\begin{equation}\label{LG2BigM}
\overline{L}_{G} \approx
\frac{m_{1}^{2}\overline{L}_{1}+m_{2}^{2}\overline{L}_{2}}{(m_{1}+m_{2})}+\frac{2m_{1}m_{2}(\overline{L}_{1,0}+\overline{L}_{2,0}+\left|\gamma)\right|}{(m_{1}+m_{2})^{2}}.
\end{equation}

{ \em Call community $Q_{j}$ a weakly symmetric community if
$\overline{L}_{j,0}=\overline{L}_{j}$, i.e. the mean distance
between the external point $x_{0}$ and the other nodes equals to
the mean distance on the entire community.}

Suppose both communities $Q_{1}$ and $Q_{2}$ are big, i.e.
$m_{1},m_{2}\gg1 $ and weakly symmetric. In this case

\begin{equation}\label{LG2BigMSimmetric}
\overline{L}_{G}=\frac{m_{1}^{2}\overline{L}_{1}+m_{2}^{2}\overline{L}_{2}+2m_{1}m_{2}(\overline{L}_{1}+\overline{L}_{2})}{(m_{1}+m_{2})}+\frac{2m_{1}m_{2}\left|\gamma)\right|}{(m_{1}+m_{2})^{2}}.
\end{equation}

If communities have the same size $m_{1}=m_{2}=m$, the same mean
distance $L=\overline{L}_{1}=\overline{L}_{2}$ and the same mean
distance to "external" nodes
$L_{0}=\overline{L}_{1,0}=\overline{L}_{2,0}$, expressions
(\ref{LG2})-(\ref{LG2BigMSimmetric})
 can be simplified by the following way

\begin{equation}\label{LG2equal}
\overline{L}_{G}=\frac{(m-1)\overline{L}}{2m-1}+\frac{2(m-1)\overline{L}_{0}}{2m-1}+\frac{m\left|\gamma\right|}{2m-1}
\end{equation}

\noindent  for two (not necessarily big) communities,

\begin{equation}\label{LG2equalBig}
\overline{L}_{G}\approx
\frac{\overline{L}}{2}+\overline{L}_{0}+\frac{\left|\gamma\right|}{2}
\end{equation}

\noindent for big communities $m \gg1 $, and

\begin{equation}\label{LG2equalBigSymmetric}
\overline{L}_{G} \approx
\frac{3}{2}\overline{L}+\frac{1}{2}\left|\gamma\right|.
\end{equation}

\noindent for weakly symmetric
($\overline{L}_{1,0}=\overline{L}_{1},\overline{L}_{2,0}=\overline{L}_{2}$)
big communities ($m \gg 1$).

\subsubsection*{Mean distance estimates for general SD-networks}

Consider a global network $G$ divided into $C$ communities $Q_{j},
j=1,...,C$ with $E$ links between communities. A macro network $M$
is obtained by replacing each community $Q_{j}$ with a single node
$g_{j}$. Any community $Q_{j}$ has $m_{j}$ nodes denoted
$q_{i,j},(i=1,...,m_j)$ and $e_{j}$ links. We assume that each
community $Q_{j}$ is connected to other communities via a single
node $q_{i0,j}$ which we call an external node. The following
notations will be used:
$\overline{L}_{j}=\frac{2\sum_{i>k=1}^{m_{j}}l(q_{i,j},q_{k,j})}{m_j(m_j-1)}$
for the mean distance of the community $Q_{j}$,
$,\overline{L}_{j,0}=\frac{\sum_{k=1}^{m_{j}}l(q_{i,j},q_{i_0,j})}{m_j-1}$
for the mean distance to the external node $q_{i_{0},j}$, and
$\overline{L}_{G}$ and $\overline{L}_{M}$ for mean distances of
the global network $G$ and the macro network $M$, respectively.

Let us repeat the previous calculations for this general case.

Again we present the mean distance $L_{G}$ as a sum of two terms
$\overline{L}_{G}=I_{1}+I_{2}$, where

\begin{widetext}

\begin{eqnarray}
I_{1}=\frac{2\sum_{j=1}^{C}\sum_{i>k=1}^{m_{j}}l(q_{i,j},q_{k,j})}{\sum_{j=1}^{C}m_{j}\left(\sum_{j=1}^{C}m_{j}-1\right)},\label{I1}\\
I_{2}=\frac{2\sum_{j>s=1}^{C}\sum_{i=1}^{m_{j}}\sum_{k=1}^{m_{s}}[l(q_{i,j},q_{i_{0},j})+l(g_{j},g_{s})+l(q_{k,s},q_{k_{0},s})]}{\sum_{j=1}^{C}m_{j}\left(\sum_{j=1}^{C}m_{j}-1\right)}=\\
\nonumber
\frac{2\sum_{j>s=1}^{C}\left[m_{s}\sum_{i=1}^{m_{j}}l(q_{i,j},q_{i_{0},j})+m_{j}\sum_{k=1}^{m_{s}}l(q_{k,s},q_{k_{0},s})\right]+2m_{j}m_{s}\sum_{j>s=1}^{C}l(g_{j},g_{s})}{\sum_{j=1}^{C}m_{j}\left(\sum_{j=1}^{C}m_{j}-1\right)}.\label{I2}
\end{eqnarray}

Recall that $\sum_{j=1}^{C}m_{j}$ represents the number $N_{G}$ of
all nodes in the global network $G$. Using the definitions for the
mean distances $\overline{L}_{j}$ of $Q_{j}$, and the mean
distances $\overline{L}_{j,0}$ to external nodes of $Q_{j}$, the
mean distance $\overline{L}_{G}$ on the global network can be
rewritten as

\begin{equation}\label{LGgen}
\overline{L}_{G}=\frac{\sum_{j=1}^{C}m_{j}(m_{j}-1)\overline{L}_{j}}{N_{G}(N_{G}-1)}+\frac{2\sum_{j>s=1}^{C}\left(m_{s}(m_{j}-1)\overline{L}_{j,0}+m_{j}(m_{s}-1)\overline{L}_{s,0}\right)}{N_{G}(N_{G}-1)}+
\frac{2\sum_{j>s=1}^{C}m_{j}m_{s}l(g_{j},g_{s})}{N_{G}(N_{G}-1)}.
\end{equation}

Let us discuss some symmetric cases and some types of possible
formal symmetries of the communities.

If all communities $Q_{j}$ are weakly symmetric communities, i.e.
$\overline{L}_{j,0}=\overline{L}_{j}$ for all $j$, then we can
replace $\overline{L}_{j,0}$ by $\overline{L}_{j}$ in
(\ref{LGgen})

\begin{equation}\label{LGgenSym}
\overline{L}_{G}=\frac{\sum_{j=1}^{C}m_{j}(m_{j}-1)\overline{L}_{j}+2\sum_{j>s=1}^{C}\left(m_{s}(m_{j}-1)\overline{L}_{j}+m_{j}(m_{s}-1)\overline{L}_{s}\right)}{N_{G}(N_{G}-1)}+
\frac{2\sum_{j>s=1}^{C}m_{j}m_{s}l(g_{j},g_{s})}{N_{G}(N_{G}-1)}.
\end{equation}
\end{widetext}

Additional simplification is possible for weakly symmetric
communities of the same size, i.e.
$\overline{L}=\overline{L}_{j}=L_{0,j}$, and $m=m_{j}$ for all $j$

\begin{equation}\label{LGgenSymEqual}
\overline{L}_{G}=\left(\frac{m-1}{Cm-1}+\frac{2(m-1)(C-1)}{Cm-1}\right)\overline{L}+\frac{m(C-1)}{Cm-1}\overline{L}_{M},
\label{eq:2}
\end{equation}

\noindent where $\overline{L}_{M}$ is the mean distance on the
macro network $M$.

If the communities are also big, i.e $m\gg 1$, then the following
asymptotic is correct

\begin{equation}\label{LGgenSymEqualBig}
\overline{L}_{G}=\left(\frac{1}{C}+2\frac{C-1}{C}\right)\overline{L}+\frac{C-1}{C}\overline{L}_{M}.
\end{equation}

If number of communities is also big $C \gg 1$ then

\begin{equation}\label{LGgenManySymEqualBig}
\overline{L}_{G}\approx2\overline{L}+\overline{L}_{M}.
\end{equation}

Because $\overline{L}\geq 1$ we have an estimate
$\overline{L}_{G}\geq 2+\overline{L}_{M}$. This inequality is
asymptotically exact for cliques \cite{social} (i.e. when $Q_{j}$
are complete graphs for any $j$). Thus in the case of dense
communities the estimate
$\overline{L}_{G}-2\geq\overline{L}_{M}$ gives an
apriori information about the macro network mean distance $L_{M}$.

\subsection*{Dilatation as an indicator of community structure existence}

Consider again a global network $G$ divided into $C$
communities with $E$ links between them. Corresponding macro
network $M$ is obtained by replacing each community $Q_{j}$ with a
single node $g_{j}$. Any community $Q_{j}$ has $m_{j}$ nodes
$q_{i,j}$ and $e_{j}$ edges (links). The following additional
notations will be used $\delta_{j}=\frac{m_{j}}{e_{j}}$,
$\delta_{G}=\frac{\sum_{j=1}^{C}m_{j}}{\sum_{j=1}^{C}e_{j}+E}$.

For SD-network it is natural to suppose that $\sum_{j=1}^{C}e_{j}\gg E$. For this case
$\delta_{G}\approx\frac{\sum_{j=1}^{C}m_{j}}{\sum_{j=1}^{C}e_{j}}$.
If all communities have the same size $m_{j}=m$ for all $j$,
$\delta_{G}\approx\frac{Cm}{\sum_{j=1}^{C}e_{j}}=\frac{C}{\sum_{j=1}^{C}\delta_{j}^{-1}}$.
If, in addition, all communities have the same density
$\delta_{j}=\delta$ for all $j$, then $\delta_{G}\approx\delta$.

These simple remarks together with estimations for the mean
distance, allow one to obtain necessary estimations for the
dilatation $D_{G}$. Thus by definition

\begin{equation}\label{definition}
D_{G}=\frac{\overline{L}_{G}}{\left[\sum_{j=1}^{C}m_{j}\right]^{\delta_{G}}},
\end{equation}

\noindent where $\overline{L}_{G}$ can be calculated using eq.
(\ref{LGgen}).

If all communities $Q_{j}$ are weakly symmetric and have the same
size $m_{j}=m$, then by equation (\ref{LGgenSymEqual}) we have

\begin{equation}\label{DSymEqual}
D_{G}\approx\frac{\left[\frac{m-1}{Cm-1}+\frac{2(m-1)(C-1)}{Cm-1}\right]\overline{L}+\frac{m(C-1)}{Cm-1}\overline{L}_{M}}{[Cm]^{\delta}}.
\end{equation}

For $m\gg 1$ we
have by equation (\ref{DSymEqual})

\begin{equation}\label{DSymEqualBig}
D_{G}\approx\frac{\left[\frac{1}{C}+2\frac{C-1}{C}\right]\overline{L}+\frac{C-1}{C}\overline{L}_{M}}{[Cm]^{\delta}}.
\end{equation}

If also $C\gg 1$ then by equation (\ref{DSymEqualBig})

\begin{equation}\label{DManySymEqualBig}
D_{G}\approx\frac{2\overline{L}+\overline{L}_{M}}{[Cm]^{\delta}}.
\end{equation}

The last asymptotic formula demonstrates that for an SD-network
with large number of similar communities the dependence of
dilatation on community type is represented by its dependence on
the network dimension $\delta$, or more accurately, on
$[Cm]^{\delta}$. For example, if communities $Q_{j}$ are complete
graphs of the same size $m$ then $\delta=\frac{2}{m-1}$ and
$[Cm]^{\frac{2}{m-1}}\rightarrow1$ for $m\rightarrow\infty$. For this theoretical case

\begin{equation}\label{DManySymEqualBigComplete}
D_{G}\approx2+\overline{L}_{M}=\overline{L}_{G},
\end{equation}

\noindent and therefore $D_{G}\geq 3$.



\begin{thebibliography}{1}
\bibitem{Newman-Girvan} M.E.J. Newman and M. Girvan, Phys. Rev. E \textbf{69}, 026113
(2004).
\bibitem{review} For recent review  see L. Danon, J. Duch, A. Arenas, and A. Diaz-Guilera,
cond-mat/0505245 (2005), and references therein.
\bibitem{social} John Scott, \textit{Social Network Analysis}, SAGE Publications, 2000.
\bibitem{Rickman} See, for example, Seppo Rickman, \textit{Quasiregular Mappings}, Springer-Verlag, 1993.
\bibitem{Guimera} R. Guimer\'{a}, M. Sales, and L. N. A. Amaral, Phys. Rev. E, \textbf{70} 025101
(2004).
\bibitem{Erdos} B. Bollobas, \textit{Random Graphs}, 2nd ed. (Cambridge University
Press, New York, 2001).
\bibitem{our-hierarchy} V. Gol'dshtein, G.A. Koganov, and G.I. Surdutovich,
cond-mat/0409298 (2004).
\bibitem{Newman-SIAM} M.E.J. Newman, SIAM Review \textbf{45}, 167 (2003).
\bibitem{actors} L. A. N. Amaral , A. Scala,  M. Barth´el´emy, and H. E. Stanley,
Proc. Natl. Acad. Sci. USA
\textbf{97}, 1114911152 (2000);  D. J. Watts and S. H. Strogatz,
Nature
\textbf{393}, 440442 (1998).
\bibitem{trainRouts} P.Sen, S. Dasgupta, A. Chatterjee, P. A. Sreeram, G. Mukherjee, and S. S. Manna, cond-mat/0208535 (2002).
\bibitem{Internet} M. Faloutsos, P. Faloutsos, and C. Faloutsos, Computer
Communications Review \textbf{29}, 251262 (1999); Q. Chen, H.
Chang, R. Govindan, S. Jamin, S. J. Shenker, and W. Willinger, in
Proceedings of the 21st Annual Joint Conference of the IEEE
Computer and Communications Societies, IEEE Computer Society
(2002).
\bibitem{CompanyDirectors} M. E. J. Newman, S. H. Strogatz, and D. J. Watts,
Phys. Rev. E \textbf{64}, 026118 (2001).
\end{thebibliography}
\end{document}